\documentclass[twoside]{article}
\usepackage{fleqn,espcrc2}
\usepackage{epsfig}

\newcommand{\AmS}{{\protect\the\textfont2
  A\kern-.1667em\lower.5ex\hbox{M}\kern-.125emS}}
\title{The topology of the Fermi surface of Bi$_2$Sr$_2$CaCu$_2$O$_{8-\delta}$ from angle resolved photoemission.}

\author{M. S. Golden, S. V. Borisenko, S. Legner, T. Pichler, C. D\"urr, M. Knupfer, J. Fink
\address{Institut f\"ur Festk\"orper- und Werkstofforschung Dresden, \\ P.O.Box 270016, D-01171 Dresden, Germany},
G. Yang, S. Abell
\address{School of Metallurgy and Materials, The University of Birmingham, \\ Birminhgam, B15 2TT, United Kingdom},
G. Reichardt\address{BESSY GmbH, \\ Albert-Einstein-Strasse 15, 12489 Berlin, Germany},
R. M\"uller, C. Janowitz\address{Humboldt-Universit\"at zu Berlin, Institut f\"ur Physik,\\ Invalidenstrasse 110, 10115 Berlin, Germany}}
       
\begin{document}

\begin{abstract}
We present a study of the topology of the normal state Fermi surface (FS) of the high T$_c$ superconductor Bi$_2$Sr$_2$CaCu$_2$O$_{8-\delta}$ (Bi2212) using angle-resolved photoemission.
We present FS mapping experiments, recorded using {\it unpolarised} radiation with high (E,{\bf k}) resolution, and an extremely dense sampling of {\bf k}-space.
In addition, synchrotron radiation-based ARPES has been used to prove the energy independence of the FS as seen by photoemission.
We resolve the current controversy regarding the normal state FS in Bi2212.
The true picture is simple, self-consistent and robust: the FS is hole-like, with the form of rounded tubes centred on the corners of the Brillouin zone. Two further types of features are also clearly observed: shadow FSs, and diffraction replicas of the main FS caused by passage of the photoelectrons through the modulated Bi-O planes.
\vspace{1pc}
\end{abstract}

\maketitle

The topology and character of the normal state Fermi surfaces of the high temperature superconductors have been the object of both intensive study and equally lively debate for almost a decade.
Angle-resolved photoemission spectroscopy (ARPES) has played a defining role in this discussion.
The pioneering work of Aebi {\it et al.} illustrated that angle-scanned photoemission using unpolarised radiation can deliver a direct, unbiased image of the complete FS of Bi2212 \cite{Aebi}, confirming the large FS centered at the corners of the Brillouin zone (BZ) predicted by band structure calculations \cite{Calculations}.
Furthermore, the use of the mapping method enabled the indentification of weak additional features ("shadow Fermi surface", SFS) which were attributed to the effects of short-range antiferromagnetic spin correlations \cite{Kampf}.
Subsequently, ARPES investigations clearly identified a further set of dispersive photoemission structures which are extrinsic and result from a diffraction of the outgoing photoelectrons as they pass through the structurally modulated Bi-O layer, which forms the cleavage surface in these systems \cite{Ding}.

Recently, this whole picture has been called into question. 
ARPES data recorded using particular photon energies (32-33 eV) have been interpreted in terms of either: a FS with missing segments \cite{Saini1}, an extra set of one dimensional states \cite{Saini2}, or an electron-like FS centred around the $\Gamma$ point \cite{Chuang}.
A further study suggests that both electron or hole-like FS pieces can be observed, depending on the photon energy used in the ARPES experiment \cite{Feng}.
These points illustrate that the situation is far from clear, and that it is essential that an unambiguous framework is arrived at for the interpretation of the ARPES data.

In this contribution, we present ARPES investigations of Bi2212, with the aim of clearing up the controversy
regarding the normal state FS topology.
We present a combination of angle-scanned photoemission data using unpolarised radiation (giving FS maps) with synchrotron-based EDCs.
In this way we can shed light on the crucial role played not only by the polarisation effects, but also by the photon energy in the photoemission data from Bi2212.
Ultimately, we are able to suggest a resolution of the current controversy, which can be shown to be a result  
of the complex, photon energy and polarisation-dependent interplay of {\it three} different types of photoemission features around the $\overline M$ point \cite{XYM}: the main FS, diffraction replicas (DRs) and the SFS.

The angle-scanned ARPES experiments were performed using monochromated, unpolarised He I radiation from a high performance source (VUV5000, Gammadata-SCIENTA) coupled to a SCIENTA SES200 analyser enabling simultaneous analysis of both the E and {\bf k}-distribution of the photoelectrons. 
The overall energy resolution was set to 30 meV and the angular resolution to $\pm$0.38 $^\circ$, which gives $\Delta${\bf k} $\leq$ 0.028 \AA$^{-1}$ (i.e. 2.4 $\%$ of $\Gamma$X).
These experiments were either carried out at room temperature or 120K.
The synchrotron-based data were recorded at 100K with $\Delta\Theta$ = $\pm$1$^\circ$ and $\Delta$E = 70 meV using a commercial 65mm goniometer-mounted analyser with radiation from the U2-FSGM beamline at the BESSY I facility.
High quality single crystals of Bi2212 \cite{Bi2212crystals} were cleaved in-situ to give mirror-like surfaces.

In Fig. 1 we show a Fermi surface map of Bi2212 with a k-point density of some 1500 EDCs per BZ quadrant.
The grey scale indicates the photoemission intensity within an energy window of 20 meV centred at the Fermi level, E$_F$.
We stress here that each of the pixels of Fig. 1 represent a 'real' EDC - no interpolation, reflection or other mathematical manipulation of the data has been carried out.
\begin{figure}[htb]
\begin{center}
\epsfig{file=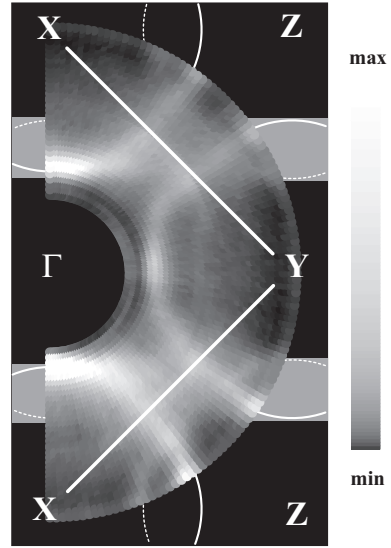,width=5cm}
\end{center}
\caption{FS map of Bi2212 taken at room temperature.}
\label{fig:largenenough}
\end{figure}

The topology of the main FS of Bi2212 for photoelectron final state energies of the order of 17 eV is evident
- Fig. 1 shows it to have the form of rounded barrels centred at the X,Y points of the BZ \cite{XYM}.
We mention at this point that the simple, almost 'traditional' topology of the FS clearly observed
here does not depend on the method used to define {\bf k}$_F$.
In addition, the point should not be overlooked that the use of unpolarised radiation in such a mapping experiment gives a relatively unbiased reproduction of the FS, without the catastrophic changes in intensity contrast which can
result from polarisation-dependent matrix element effects. 
Also evident in Fig. 1 is the presence of the shadow FS, which is particularly clear  in the lower right
corner of the map (the dotted lines show the form of the SFS).

Nevertheless, the question remains as to the validity of this FS topology when 'seen'
using photoelectrons with higher final state energies.
It could be argued that the final states (17-20 eV above
E$_F$) accessed with 'traditional' photon energies are not sufficiently high to guarantee 
free electron-like final states. 
In particular, it is the exact situation around the $\overline M$ point which is central to the debate, as
it is in this region of {\bf k}-space where the 'closing' of the main FS arcs to give a $\Gamma$-centred
(electron-like) FS has been proposed. 
On using photon energies between 32 - 33eV (giving a final state energy of around 28eV) a number of groups have concluded a radically different topology for the normal state FS of Bi2212 \cite{Chuang,Feng}.
In order to address this question, we have measured EDCs of Bi2212 using sychrotron radiation of different energies.
Part of these data are shown in Fig. 2, which displays two series of EDCs recorded in the normal state (T=100K)
along the $\Gamma$$\overline M$Z line in {\bf k}-space.

\begin{figure}[htb]
\begin{center}
\epsfig{file=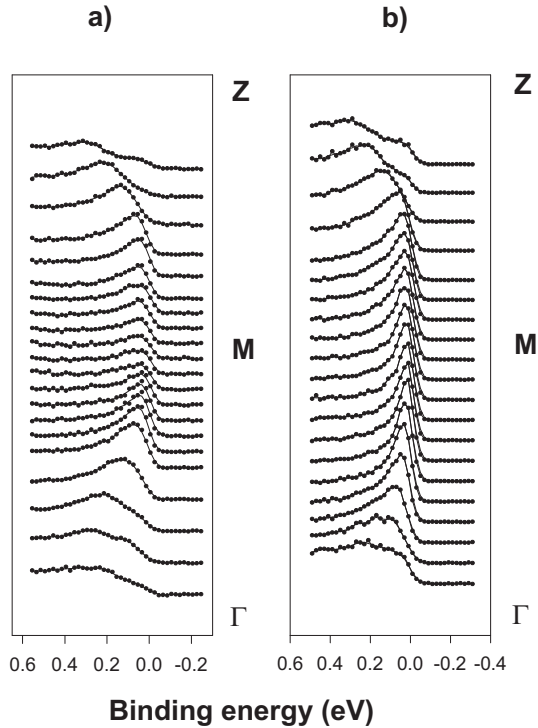,width=7cm}
\end{center}
\caption{EDCs of Bi2212 along $\Gamma$$\overline M$Z taken with a) 32eV and b) 50eV photon energies. 
The spectra are linearly offset in {\bf k}-space.
}
\end{figure}

The data with h$\nu$=32 eV are very similar to those reported in \cite{Chuang} and are recorded in the same
experimental geometry. 
It is evident that the spectral weight of the states related to the extended saddle point singularity 
is strongly reduced around the $\overline M$ point, in agreement with recent theoretical calculations \cite{Bansil}.
This reduction could indeed be seen as a sign of a FS crossing, followed by a reappearance
of the band between $\overline M$Z.
However, taking a photon energy of 50eV, for which no-one would doubt the validity of a quasi-free electron-like final state, the picture is completely different.
These data look like those recorded for h$\nu$=20-25eV clearly showing the approach of the band to the E$_F$ but giving no evidence of a Fermi level crossing - i.e. the extended saddle point singularity scenario is confirmed.

The singular nature of the data for h$\nu$$\sim$32-33 eV, combined with the reappearance of the 'traditional' FS
picture for higher photon energies is a strong indication that the former are the result of strong matrix
element-derived suppression of the spectral weight of the main band and not a FS crossing.

This explanation is confirmed by the data shown in Fig. 3, 
in which we present a detailed momentum map of the normal state FS of Bi2212 recorded at 120 K
in order to capture more sharply the details around the $\overline M$-point.
\begin{figure}[htb]
\begin{center}
\epsfig{file=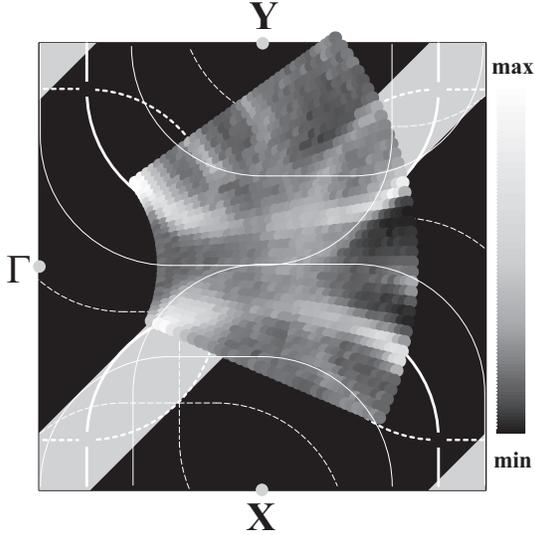,width=7cm}
\end{center}
\caption{FS map of Bi2212 taken at 120K. The lines denote the main FS (thick solid), SFS (thick dashed) and the 
1st (thin solid) and 2nd (thin dashed) order DRs.
The image is based upon 1300 EDCs \cite{Normalization}.}
\end{figure}
Fig. 3 clearly shows the richness of structure in the ARPES data around $\overline M$. This arises from the complex interplay between main FS, DRs and the SFS features. As an example of this, one can see the overlap of the SFS and 1st order DR features at ca. 0.6($\pi$,$\pi$) as a bright spot on the map.
We emphasize that only an analysis of uninterpolated data recorded with high (E,{\bf k})-resolution on an extremely fine {\bf k}-mesh can enable the discrimination between the numerous features concentrated within this small region of the BZ. Furthermore, 
the DRs of the main FS also lead to a bundling of intensity along a ribbon centered on the (0,-$\pi$)-($\pi$,0) line - indicated by grey shading in Figs. 1 and 3. 
Indeed, our data offer a natural explanation for the observations made with h$\nu$=32-33eV. 
The matrix-element mediated reduction of the saddle point intensity for these photon energies (Fig. 2 and \cite{Bansil})
explains both the suppression of spectral weight directly along the (0,-$\pi$)-($\pi$,0) cut observed in \cite{Saini1},
and means that the edges of the ribbon will become relatively more intense.
Thus, crossing of the two edges of the ribbon feature could be mistakenly interpreted as a 'main'
band crossing along the $\Gamma$-$\overline M$ line \cite{Chuang,Feng}.

In conclusion, we have shown that high (E,{\bf k}) resolution, high {\bf k}-density angle-scanned photoemission data-sets combining the advantages of both the mapping and EDC approaches give a self-consistent and robust picture of the nature and topology of the FS in Bi2212. From the data presented here and 
from a comparison of pristine and Pb-doped Bi2212 \cite{us} it is clear that three different features 
are present in the ARPES data of Bi2212. These are a hole-like main FS with the topology of a curved barrel centred around the X,Y points, extrinsic DRs of the main FS due to the Bi-O modulation which lead to high intensity ribbons centered on the (0,-$\pi$)-($\pi$,0) line, and a shadow FS. 
The recent controversy as regards the FS topology most likely resulted from 
the use of unfavorable experimental conditions.

We gratefully acknowledge fruitful discussions with H. Eschrig and D. van der Marel. Part of this work has been supported by the BMBF (05 SB8BDA 6), the DFG (Graduiertenkolleg 'Struktur- und Korrelationseffekte in Festk\"orpern' der TU-Dresden) and the SMWK (4-7531.50-040-823-99/6).


\begin{thebibliography}{9}
\bibitem{Normalization} Each EDC is normalised by division with the total integrated intensity of the spectrum. This gives a new criterion for determining {\bf k}$_{F}$, and can be viewed as a combination of the $I_{max}$ and $\nabla$n({\bf k}) methods.
\end{thebibliography}
\end{document}